\documentclass[epj]{svjour}
\usepackage{graphics,amssymb}
\def\openone{\leavevmode\hbox{\small1\kern-3.8pt\normalsize1}}
\def\mf{{\mbox{\tiny\em MFA}}}

\begin{document}
\title{
Phase diagram of neutron star quark matter in nonlocal chiral models}
\author{D. G\'omez Dumm\inst{1,2}, D.B. Blaschke\inst{3,4,5},
A.G. Grunfeld\inst{6}, T. Kl\"ahn\inst{4,7} \and N.N. Scoccola\inst{1,6,8}
% etc
}                     % Do not remove
\institute{CONICET, Rivadavia 1917, 1033 Buenos Aires, Argentina
\and Dpto.\ de F\'{\i}sica, Universidad Nacional de La Plata, C.C.\ 67,
1900 La Plata, Argentina
\and
Institut f\"ur Physik, Universit\"at Rostock, Universit\"atsplatz 3,
D-18051 Rostock, Germany
\and
Instytut Fizyki Teoretycznej, Uniwersytet Wroc{\l}awski, pl. M. Borna 9,
50-204 Wroc{\l}aw, Poland
\and
Bogoliubov  Laboratory of Theoretical Physics, JINR Dubna, Joliot-Curie
Street 6, 141980  Dubna, Russia
\and
Physics Department, Comisi\'on Nacional de Energ\'{\i}a At\'omica,
Av.\ Libertador 8250, 1429 Buenos Aires, Argentina
\and
Gesellschaft f\"ur Schwerionenforschung mbH, Planckstr. 1,
D-64291 Darmstadt, Germany
\and Universidad Favaloro, Sol{\'\i}s 453, 1078 Buenos Aires, Argentina}
\date{Received: date / Revised version: date}
\abstract{We analyze the phase diagram of two-flavor quark matter under
neutron star constraints for a nonlocal covariant quark model within the
mean field approximation. Applications to cold compact stars are
discussed.
} %end of abstract

\authorrunning{D. G\'omez Dumm {\em et al.}}

\titlerunning{Phase diagram of neutron star quark matter in nonlocal chiral
models}

\maketitle
\section{Introduction}
\label{intro}

The characteristics of the QCD phase diagram is presently an important
open subject of research in particle physics. In particular, the behavior
of strongly interacting matter in the region of low temperatures and large
baryon densities could be tested against observational constraints from
neutron stars~\cite{Blaschke:2001uj}. Since in this thermodynamical region
one finds strong difficulties when trying to perform lattice QCD
calculations, the pictures emerging from different effective models of
strong interactions deserve great interest from the theoretical point of
view. Here we focus on a two-flavor chiral quark model which includes
covariant nonlocal four-fermion interactions, motivated by an effective
one-gluon exchange (OGE) picture. Nonlocality arises naturally in the
context of several successful approaches to low energy quark dynamics,
such as the instanton liquid model~\cite{SS98} and the Schwinger-Dyson
resummation techniques~\cite{RW94}, and it is also a well-known feature of
lattice QCD~\cite{SLW01}.

\section{Formalism}

The Euclidean action for the nonlocal model considered here, in the case
of two light flavors and anti-triplet diquark interactions, is given by
\begin{eqnarray}
S_E & = & \int d^4 x \ \bigg\{ \bar \psi (x) \left(- i \rlap/\partial
+ m \right) \psi (x) \nonumber \\
& & - \frac{G}{2} j^f_M(x) j^f_M(x) -
\frac{H}{2} \left[j^a_D(x)\right]^\dagger j^a_D(x) \bigg\} \,.
\label{action}
\end{eqnarray}
Here $m$ is the current quark mass, which is assumed to be equal for $u$
and $d$ quarks, and $j_{M,D}$ are mesonic and diquark nonlocal currents.
The nonlocality is introduced here in a covariant way, through a separable
interaction arising from an effective OGE picture. In this way, we have
\begin{eqnarray}
j^f_M (x) &=& \int d^4 z \  g(z) \
\bar \psi(x+\frac{z}{2}) \ \Gamma_f \ \psi(x-\frac{z}{2})\,,
\nonumber \\
j^a_D (x) &=&  \int d^4 z \ g(z)\, \bar \psi_c(x+\frac{z}{2}) \
i \gamma_5 \tau_2 \lambda_a \ \psi(x-\frac{z}{2}) \ ,
\label{cuOGE}
\end{eqnarray}
where $\psi_c(x) = \gamma_2\gamma_4 \,\bar \psi^T(x)$, $\Gamma_f = (
\openone, i \gamma_5 \vec \tau )$, and $\vec \tau$ and $\lambda_a$, with
$a=2,5,7$, stand for Pauli and Gell-Mann matrices acting on flavor and
color spaces, respectively. The function $g(z)$ is a form factor
that characterizes the  nonlocal interaction.

The effective action in Eq.~(\ref{action}) might arise via Fierz
rearrangement from some underlying more fundamental interactions, and is
understood to be used ---at the mean field level--- in the Hartree
approximation. In general, the ratio of coupling constants $H/G$ would be
determined by these microscopic couplings; for example, OGE interactions
lead to $H/G=0.75$. Since the precise derivation of the effective
couplings from QCD is not known, here we will leave $H/G$ as a free
parameter.

Standard bosonization of the theory leads, in the mean field
approximation, to a thermodynamical potential per unit volume given by
\begin{eqnarray}
\Omega^\mf & = & \frac{ \bar \sigma^2 }{2 G} + \frac{|\bar
\Delta|^2}{2 H} - \frac{T}{2}\!\sum_{n=-\infty}^{\infty} \int
\frac{d^3 \vec{p}}{(2\pi)^3} \, \ln \mbox{det} \left[
\frac{S^{-1}}{T} \right] , \ \ \ \ \ \label{z}
\end{eqnarray}
where the inverse propagator $S^{-1}(\bar \sigma,\bar \Delta)$ is a $48
\times 48$ matrix in Dirac, flavor, color and Nambu-Gorkov spaces (its
explicit form is given in Ref.~\cite{GomezDumm:2005hy}). Here $\bar\sigma$
and $\bar\Delta$ stand for the mean field values of scalar meson and
diquark fields. Owing to the nonlocality, they come together with
momentum-dependent form factors. The values of $\bar\sigma$ and
$\bar\Delta$ can be obtained from the coupled gap equations
\begin{equation}
\frac{ d \Omega^\mf}{d\bar \Delta} \ = \ 0 \ ,
\hspace{1cm} \rule{0cm}{.7cm}
\frac{ d \Omega^\mf}{d\bar \sigma} \ = \ 0 \ .
\label{sigud}
\end{equation}

In general one has to consider a different chemical potential $\mu_{fc}$
for each quark flavor $f$ and color $c$. However, when the system is in
chemical equilibrium, not all $\mu_{fc}$ are independent. In our case, it
can be seen~\cite{GomezDumm:2005hy} that they can be written in terms of
only three quantities: the baryon chemical potential $\mu_B$, the quark
electric chemical potential $\mu_{Q}$ and the color chemical potential
$\mu_8$. Defining $\mu=\mu_B/3$, the corresponding relations read
\begin{eqnarray}
\mu_{qr} = \mu_{qg} &=& \mu + Q_q\; \mu_{Q} + \mu_8 /3 \ , \nonumber \\
\mu_{qb} &=& \mu + Q_q\; \mu_{Q} - 2\,\mu_8/3 \ ,
\label{chemical}
\end{eqnarray}
where, $q=u,d$, and $Q_q$ are quark electric charges.

In the core of neutron stars, in addition to quark matter we have
electrons. The latter can be thermodynamically treated as a free Fermi
gas, and their contribution has to be added to the grand canonical
thermodynamical potential. Moreover, quark matter and electrons have to be
in $\beta$- equilibrium. Thus, assuming that antineutrinos escape from the
stellar core, we must have
\begin{equation}
\mu_{dc} - \mu_{uc} = - \mu_{Q} = \mu_e \ .
\label{beta}
\end{equation}

If we now require the system to be electric and color charge neutral, the
number of independent chemical potentials reduces further: $\mu_e$ and
$\mu_8$ are fixed by the conditions of vanishing electric and color
densities. In this way, for each value of $T$ and $\mu$ we should find the
values of $\bar\Delta$, $\bar\sigma$, $\mu_e$ and $\mu_8$ that solve
Eqs.~(\ref{sigud}), supplemented by the $\beta$-equilibrium and electric
and charge neutrality conditions.

\section{Numerical results}

According to previous analyses carried out within nonlocal
scenarios~\cite{DGS04}, the results are not expected to show a strong
qualitative dependence on the shape of the form factor. Thus we will
consider here a simple, well-behaved Gaussian function
\begin{eqnarray}
g(p^2) = \exp(-p^2/\Lambda^2) \ ,
\label{reg2}
\end{eqnarray}
where $\Lambda$ is a free parameter, playing the r\^ole of an ultraviolet
cut-off.

For definiteness, for the input parameters we choose $m = 5.12$ MeV,
$\Lambda = 827$~MeV and $G \Lambda^2= 18.78$. These values are fixed so as
to reproduce the empirical values for the pion mass $m_\pi$ and decay
constant $f_\pi$, and lead to a phenomenologically reasonable value for
the chiral condensate~\cite{nos2} at vanishing $T$ and $\mu_B$, namely
$\langle 0|\bar q q|0\rangle^{1/3} = - 250$~MeV. The only remaining free
parameter is the coupling strength $H$ in the scalar diquark channel. We
choose here values for $H/G$ in the range from 0.5 to 1, i.e.\ around the
Fierz value $H/G=0.75$ discussed above.
%%%%%%%%%%%
\begin{figure}[ht]
%\hspace*{.2cm}       % Give the correct figure height in cm
\resizebox{0.479\textwidth}{!}{%
  \includegraphics{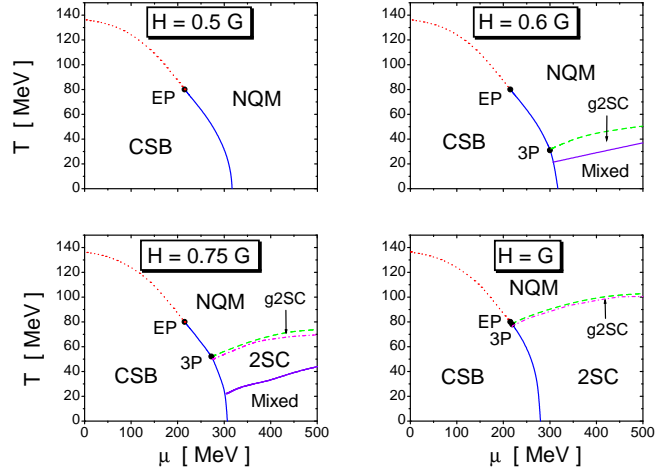}
} \caption{Phase diagrams for the nonlocal OGE-based model, for different
values of the ratio $H/G$, under neutron star constraints.}
\label{fases}       % Give a unique label
\end{figure}

Our results for the phase diagrams are shown in Fig.~\ref{fases}, where we
plot the phase transition curves on $T-\mu$ diagrams for different ratios
$H/G$. In the graphs we show the regions corresponding to different
phases, as well as the position of triple points (3P) and end points (EP).
Besides the region of low $T$ and $\mu$, in which the chiral symmetry is
broken (CSB), one finds normal quark matter (NQM) and two-flavor
superconducting (2SC) phases. Between CSB and NQM phases one has first
order and crossover transitions ---represented by solid and dotted lines,
respectively---, whereas between NQM and 2SC regions, in all cases, we
find a second order phase transition ---dashed lines in the diagrams of
Fig.~\ref{fases}. Close to this NQM/2SC phase border, the dashed-dotted
lines in the graphs delimit a band that corresponds to the so-called
gapless 2SC (g2SC) phase. In this region, in addition to the two gapless
modes corresponding to the unpaired blue quarks, the presence of flavor
asymmetric chemical potentials gives rise to another two gapless fermionic
quasiparticles. For the range of parameters considered here, however, the
g2SC region is too narrow to lead to sizeable effects. Finally, for
intermediate values of $H/G$ we find a region in which a certain volume
fraction of the quark matter undergoes a transition to the 2SC phase
coexisting with the remaining NQM phase. This is a 2SC-NQM mixed phase in
which the system realizes the constraint of electric neutrality globally:
the coexisting phases have opposite electric charges which neutralize each
other, at a common equilibrium pressure.

It can be seen that the 2SC phase region becomes larger when the ratio
$H/G$ is increased. This is not surprising, since $H$ is the effective
coupling governing the quark-quark interaction that gives rise to the
pairing. As a general conclusion, it can be stated that even under neutron
star constraints, provided the ratio $H/G$ is not too low, the nonlocal
scheme favors the existence of color superconducting phases at low
temperatures and moderate chemical potentials (we do not find 2SC only for
$H/G = 0.5$). This is in contrast to the situation in the NJL
model~\cite{Bub04}, where the existence of a 2SC phase turns out to be
rather dependent on the input parameters. Our results are also
qualitatively different from those obtained in the case of noncovariant
nonlocal models~\cite{Aguilera:2004ag}, where above the chiral phase
transition the NQM phase is preferable for values of the coupling ratio
$H/G\lesssim 0.75$, and a color superconducting phase can be found only for
$H/G\approx 1$.
It is now desirable to extend the present studies to other pairing
channels relevant for neutron star cooling, such as spin-1 pairing, where
at present only the NJL~\cite{Aguilera:2005tg} and noncovariant nonlocal
models~\cite{Aguilera:2006cj} have been considered.

\section{Application to cold compact stars}

One of the most important present applications of the microscopic
approaches to quark matter is to study whether such a state of matter can
exist in the interior of cold compact stars. Let us briefly discuss our
ongoing investigations on this issue. Unfortunately, a consistent
relativistic approach to the quark-hadron phase transition, where hadrons
appear as bound states of quarks, has not been developed up to now. Thus,
we apply a so-called two-phase description, in which the nuclear matter
phase is described within the relativistic Dirac-Brueckner-Hartree-Fock
(DBHF) approach (see e.g.~Ref.~\cite{Klahn:2006ir}) and the transition to a
quark matter phase is obtained by a Maxwell construction. From the curves
presented in the previous section, it can be seen that in our OGE-inspired
nonlocal model one finds a relatively low value of the critical density at
T=0, hence some extra repulsion is needed in order to obtain a more
realistic value. We have found that this can be achieved by including some
small additional interaction in the omega vector meson channel, which does
not affect in general the qualitative features of the phase diagrams
discussed in the previous section. Given the corresponding equation of
state, the mass and structure of spherical, nonrotating stars
%(to which we have limited ourselves so far)
is obtained by solving the Tolman-Oppenheimer-Volkov equations.
Our preliminary results confirm the findings of Ref.~\cite{Klahn:2006iw},
i.e.~that compact stars with quark matter cores are consistent with modern
observations. Moreover, for the nonlocal
quark models described here, this is found to be possible for values of
$H/G$ closer to the standard, OGE motivated ratio $H/G=0.75$.

\section{Conclusions}

We have studied the phase diagram of two-flavor quark matter under neutron
star constraints for a nonlocal, covariant quark model within the mean
field approximation. The form of the nonlocal coupling has been motivated
by a separable approximation of the OGE interaction. We have considered a
nonlocal form factor of a Gaussian shape, and the model parameters
(current quark mass $m$, coupling strength $G$, UV cutoff $\Lambda$) have
been fixed so as to obtain adequate values for the pion mass, the pion
decay constant and the chiral condensate at vanishing $T$ and $\mu_B$.

After the numerical evaluation of the gap equations at finite temperature
and chemical potential, considering different values for the coupling
strength in the scalar diquark channel, we have found that different
low-tem\-per\-a\-ture quark matter phases can occur at intermediate
densities: normal quark matter (NQM), pure superconducting (2SC) quark
matter and mixed 2SC-NQM phases. A band of gapless 2SC phase appears at
the border of the superconducting region, but this occurs in general at
nonzero temperatures and should not represent a robust feature for compact
star applications. Finally, in the context of the nonlocal theory
discussed here, we have obtained preliminary results showing that compact
stars with quark matter cores turn out to be consistent with modern
observations.

This work has been supported in part by CONICET and ANPCyT (Argentina),
under grants PIP 6009, PIP 6084, PICT02-03-10718 and PICT04-03-25374, and
by a scientist exchange program between Germany and Argentina funded
jointly by DAAD and ANTORCHAS under grants No. DE/04/27956 and 4248-6,
respectively. T.K. acknowledges support by the Virtual Institute
VH-VI-041 of the Helmholtz Association and by the GSI Darmstadt.


\begin{thebibliography}{}
\bibitem{Blaschke:2001uj}
D.~Blaschke, and D.~Sedrakian (Eds.),
{\it Superdense QCD matter and compact stars},
(NATO Science Series {\bf II/197}, Springer 2006).

\bibitem{SS98}
T. Sch\"afer and E.V. Shuryak, Rev. Mod. Phys. {\bf 70} (1998)
323.

\bibitem{RW94}
C.D.~Roberts and A.G.~Williams, Prog.\ Part.\ Nucl.\ Phys.\ {\bf 33}
(1994) 477;
C.~D.~Roberts and S.~M.~Schmidt, Prog.\ Part.\ Nucl.\ Phys.\ {\bf 45}
(2000) S1.

\bibitem{SLW01}
J.~Skullerud, D.B.~Leinweber and A.G.~Williams,
Phys.\ Rev.\ D {\bf 64} (2001) 074508.

\bibitem{GomezDumm:2005hy}
  D.~Gomez Dumm, D.~B.~Blaschke, A.~G.~Grunfeld and N.~N.~Scoccola,
  %``Phase diagram of neutral quark matter in nonlocal chiral quark models,''
  Phys.\ Rev.\  D {\bf 73} (2006) 114019.
%  [arXiv:hep-ph/0512218].
%%CITATION = HEP-PH 0512218;%%

\bibitem{DGS04}
R.~S.~Duhau, A.~G.~Grunfeld and N.~N.~Scoccola,
Phys.\ Rev.\ D {\bf 70} (2004) 074026.

\bibitem{nos2}
D.~G\'omez Dumm, A.~G.~Grunfeld and N.~N.~Scoccola,
Phys.\ Rev.\ D {\bf 74} (2006) 054026;
D.~G\'omez Dumm and N.~N.~Scoccola, Phys.\ Rev.\ C {\bf 72} (2005) 014909.

\bibitem{Bub04}
M.~Buballa, Phys.\ Rept.\ {\bf 407} (2005) 205.

\bibitem{Aguilera:2004ag}
D.~N.~Aguilera, D.~Blaschke and H.~Grigorian,
Nucl.\ Phys.\  A{\bf 757} (2005) 527.
%  [arXiv:hep-ph/0412266].
%%CITATION = HEP-PH 0412266;%%

\bibitem{Aguilera:2005tg}
  D.~N.~Aguilera, D.~Blaschke, M.~Buballa and V.~L.~Yudichev,
  %``Color-spin locking phase in two-flavor quark matter for compact star
  %phenomenology,''
  Phys.\ Rev.\ D {\bf 72} (2005) 034008.
%  [arXiv:hep-ph/0503288].
%%CITATION = HEP-PH 0503288;%%

\bibitem{Aguilera:2006cj}
  D.~N.~Aguilera, D.~Blaschke, H.~Grigorian and N.~N.~Scoccola,
  %``Nonlocality effects on color spin locking condensates,''
  Phys.\ Rev.\  D {\bf} (in press), arXiv:hep-ph/0604196.
%%CITATION = HEP-PH 0604196;%%

\bibitem{Klahn:2006ir}
T. Kl\"ahn {\em et al.},
Phys.\  Rev.\  C {\bf 74} (2006) 035802.
%  [arXiv:nucl-th/0602038].
  %%CITATION = NUCL-TH 0602038;%%

\bibitem{Klahn:2006iw}
T. Kl\"ahn {\em et al.},
arXiv:nucl-th/0609067.

\end{thebibliography}
\end{document}